\documentclass[10pt, conference]{IEEEtran}
\IEEEoverridecommandlockouts
\usepackage{cite}
\usepackage{amsmath,amssymb,amsfonts}
\usepackage{algorithmic}
\usepackage{graphicx}
\usepackage{textcomp}
\usepackage{xcolor}
\usepackage{caption}
\usepackage{subcaption}
\usepackage[hyphens]{url} 
\def\BibTeX{{\rm B\kern-.05em{\sc i\kern-.025em b}\kern-.08em
		T\kern-.1667em\lower.7ex\hbox{E}\kern-.125emX}}
\begin{document}
	
\title{Forecasting SQL Query Cost at Twitter}

\author{\IEEEauthorblockN{Chunxu Tang, Beinan Wang, Zhenxiao Luo, Huijun Wu, Shajan Dasan, Maosong Fu, Yao Li, Mainak Ghosh, \\Ruchin Kabra, Nikhil Kantibhai Navadiya, Da Cheng, Fred Dai, Vrushali Channapattan, and Prachi Mishra}
	\IEEEauthorblockA{\textit{Twitter, Inc.} \\
		San Francisco, USA \\
		\{chunxut, beinanw, zluo, huijunw, sdasan, mfu, yaoli, mghosh, rkabra, nnavadiya, dac, fdai, vrushali, prachim\}@twitter.com
	}
}
\maketitle

\begin{abstract}
	With the advent of the Big Data era, it is usually computationally expensive to calculate the resource usages of a SQL query with traditional DBMS approaches. Can we estimate the cost of each query more efficiently without any computation in a SQL engine kernel? Can machine learning techniques help to estimate SQL query resource utilization? The answers are yes. We propose a SQL query cost predictor service, which employs machine learning techniques to train models from historical query request logs and rapidly forecasts the CPU and memory resource usages of online queries without any computation in a SQL engine. At Twitter, infrastructure engineers are maintaining a large-scale SQL federation system across on-premises and cloud data centers for serving ad-hoc queries. The proposed service can help to improve query scheduling by relieving the issue of imbalanced online analytical processing (OLAP) workloads in the SQL engine clusters. It can also assist in enabling preemptive scaling. Additionally, the proposed approach uses plain SQL statements for the model training and online prediction, indicating it is both hardware and software-agnostic. The method can be generalized to broader SQL systems and heterogeneous environments. The models can achieve 97.9\% accuracy for CPU usage prediction and 97\% accuracy for memory usage prediction.
\end{abstract}


\section{Introduction}\label{sec.intro}

In the data platform at Twitter, there is a large effort in pursuing high scalability and availability to fulfill the increasing needs for data analytics on the sea of data. Twitter runs multiple large Hadoop clusters with over 300PB of data \cite{twitter-hadoop}, which are among the biggest in the world. To overcome the performance issues in developing and maintaining SQL systems with increasing volumes of data, we designed a large-scale SQL federation system across on-premises and cloud Hadoop clusters, paving the path for democratizing data analytics and improving productivity at Twitter. The SQL federation system is processing around 10PB of data daily.

During the daily operation of large-scale SQL systems, we found that the lack of forecasting SQL query resource usages is problematic. First of all, data system customers would like to know the resource consumption estimation of their queries. We received complaints that they wasted a considerable amount of time waiting for ad-hoc queries to complete but finally canceled them in frustration\footnote{From an analysis of query request logs in a three-month session, 16.2\% of uncompleted queries are related to user cancellation.}. We found that there is a correlation between the CPU time and wall clock time of ad-hoc queries in Twitter OLAP workload. Once the query resource is predicted, users can know approximately how long they will wait. Second, according to the convoy effect, query scheduling requires an estimate of the immediate workload in the SQL system. Without proper query scheduling, the cluster can be overwhelmed because of resource-consuming queries which can easily occupy most resources in a cluster in as short as 10s. Third, elastic scaling needs query resource usage forecasting. Due to the rapid impact of resource-consuming queries, a SQL system has to scale ahead of the actual processing of these queries.

A query cost prediction system can provide the following benefits for large-scale SQL systems:

\begin{itemize}
	\item \textbf{Rapid estimate of CPU and memory usage of a query for customers.} They will have an intuitive idea of resources a query may consume and an anticipated billing associated.
	\item \textbf{Improved query scheduling.} With the forecasted resource usage of a query, we can differentiate whether this is a resource-consuming query or a lightweight query. This will help balance the workloads on different clusters, before executing any queries in a SQL engine kernel.
	\item \textbf{Enabling preemptive scaling.} Because resource-intensive queries can occupy most resources of a cluster in a short time, the prediction of resource usages can help to trigger preemptive scaling if it alerts that the immediate workload may overwhelm a preset configurable threshold. 
\end{itemize}

As the query cost needs to be predicted before the query is executed in a SQL engine, we do not leverage traditional DBMS approaches such as the cost model \cite{wu2013predicting, leis2015good, baldacci2018cost}. By contrast, we propose to utilize machine learning techniques to train two models from historical SQL query request logs for CPU time and peak memory prediction. We categorize the queries into various ranges according to their CPU time and peak memory. Unlike some prior work \cite{gupta2008pqr,ganapathi2009predicting, akdere2012learning} of generating features from query plans, we extract features from plain SQL statements. Our evaluation results show that the XGBoost classifier together with the TF-IDF vectorization achieves high performance regarding accuracy as well as precision and recall for each category.

This paper makes the following contributions:

\begin{enumerate}
	\item Motivated by challenges that customers and engineers are facing in a large-scale SQL system, we introduce a SQL query cost prediction system for resource usage estimation independent from SQL engines.
	\item We harness supervised machine learning techniques to forecast both CPU and memory resource usages of individual SQL queries. As the proposed approach does not have any specific hardware or software requirements, it can be generalized to broader SQL systems and heterogeneous environments. Moreover, from our observation, the P95 (95th percentile) of the query planning latency in the SQL federation system is 9s, and the latency can be as high as more than 30s. By contrast, the proposed method only requires a constant time for the prediction (around 200ms) for all types of SQL queries.
	\item We deployed the trained models in the online production environment for resource usage prediction. We observed that the model accuracy may drop to around 92\% in four weeks and verified the concept drift. Moreover, we implemented a quantitative analysis on the monitoring results.
\end{enumerate}

The remainder of this paper is organized as follows. We discuss related work in Section \ref{sec:related-work}, describe the design of the SQL federation system in Section \ref{sec:federation}, demonstrate the architecture of the query cost prediction system in Section \ref{sec:query-predictor-overview}, explain the data preprocessing in Section \ref{sec:preprocessing}, and present training and evaluation in Section \ref{sec:training}. Section \ref{sec.conclusion} concludes the paper.

\section{Related Work}\label{sec:related-work}

With the increasing volume of data, many distributed SQL engines, targeting analyzing Big Data, emerged in the recent decade. For example, Apache Hive \cite{thusoo2009hive} is a data warehouse built on top of Hadoop, providing a SQL-like interface for data querying. Spark SQL \cite{armbrust2015spark} is a module integrated with Apache Spark, powering relational processing to Spark data structures. Presto \cite{sethi2019presto} is a distributed SQL engine, targeting ``SQL on everything''. It can query data from multiple sources which is a major advantage. At the same time, there are some commercial products such as Google BigQuery \cite{sato2012inside} (a public implementation of Dremel \cite{melnik2010dremel, melnik2020dremel}), Amazon Redshift \cite{gupta2015amazon}, and Snowflake \cite{dageville2016snowflake}, acting as fully-managed cloud data warehouses providing out-of-box data analytics experience. 

Some prior work pioneers the study of resource prediction and scaling in a database system. For example, Narayanan et al. \cite{narayanan2005continuous} proposed a Resource Advisor to answer ``what-if'' questions about resource upgrades to predict the buffer pool size on online transaction processing (OLTP) workloads. Rogers et al. \cite{rogers2010generic} described a ``white-box'' and a ``black-box'' approach to build a resource provisioning framework on top of an Infrastructure-As-A-Service cloud. Similarly,  Das et al. \cite{das2016automated} presented an automated demand-driven resource scaling approach by deriving a set of robust signals from the database engine and combining them for better scaling.

Recently, some researchers began to utilize machine learning techniques to solve resource prediction problems. Some work concentrated on optimized planning through identifying and forecasting workload patterns. For example, Mozafari et al. \cite{mozafari2013performance} developed the DBSeer system, employing statistical models for resource and performance prediction. Ma et al. \cite{ma2018query} created a robust forecasting framework to predict the expected arrival rate of queries based on historical data. They studied Linear Regression, Recurrent Neural Network, and Kernel Regression. Higginson et al. \cite{higginson2020database} utilized time series analysis and supervised machine learning to identify traits for database workload capacity planning.

Some other work focused on predicting metrics of a query, such as query performance prediction, usually with the help of query plans. For example, Gupta et al. \cite{gupta2008pqr} introduced a classification tree structure obtained from historical data of queries with the query plan to predict the execution time of a query. Ganapathi et al. \cite{ganapathi2009predicting} built machine learning models with query plan feature vectors to predict metrics such as elapsed time, disk I/Os, and message bytes. Akdere et al. \cite{akdere2012learning} evaluated predictive modeling techniques, ranging from plan-level models to operator-level models. Marcus et al. \cite{marcus2019plan} introduced a plan-structured neural network to predict query performance. Some work, such as \cite{duggan2011performance}, \cite{duggan2014contender}, and \cite{venkataraman2016ernest}, also extended the prediction to concurrent query performance prediction.
 
The existing plan-based approaches rely on query planning in SQL engines, indicating limitations under the scenario of predicting query resource usages for query scheduling and preemptive scaling when SQL engines are not involved. To resolve this problem, we propose a machine learning approach learning from plain SQL statements' features to forecast both CPU time and peak memory without dependency on any SQL engines or query plans, fulfilling the requirements of query scheduling and preemptive scaling.

\section{SQL Federation System}\label{sec:federation}

\begin{figure}[htb]
	\centering
	\includegraphics[scale=0.58]{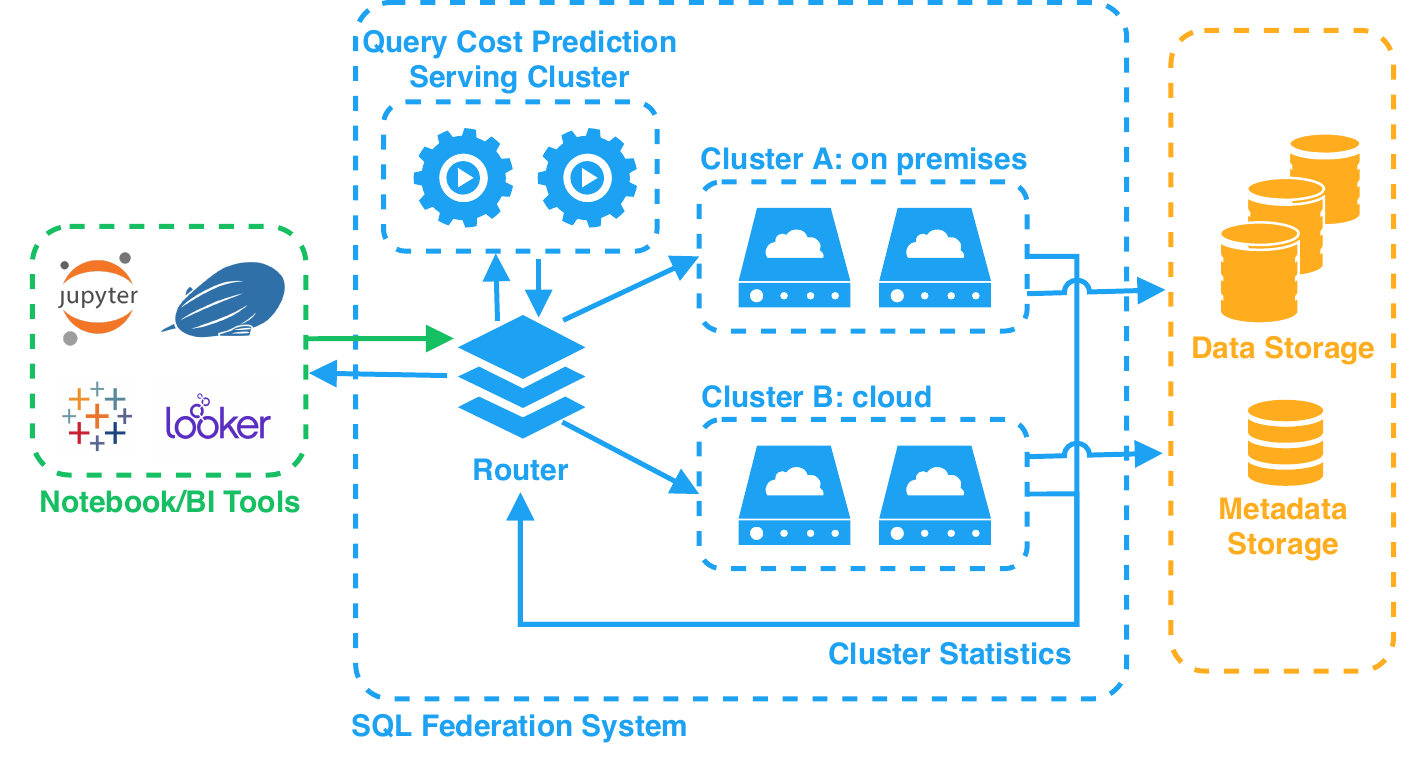}
	\caption{SQL federation system at Twitter.}
	\label{fig:sql-system}
\end{figure}

\begin{figure*}[htb]
	\centering
	\includegraphics[scale=0.63]{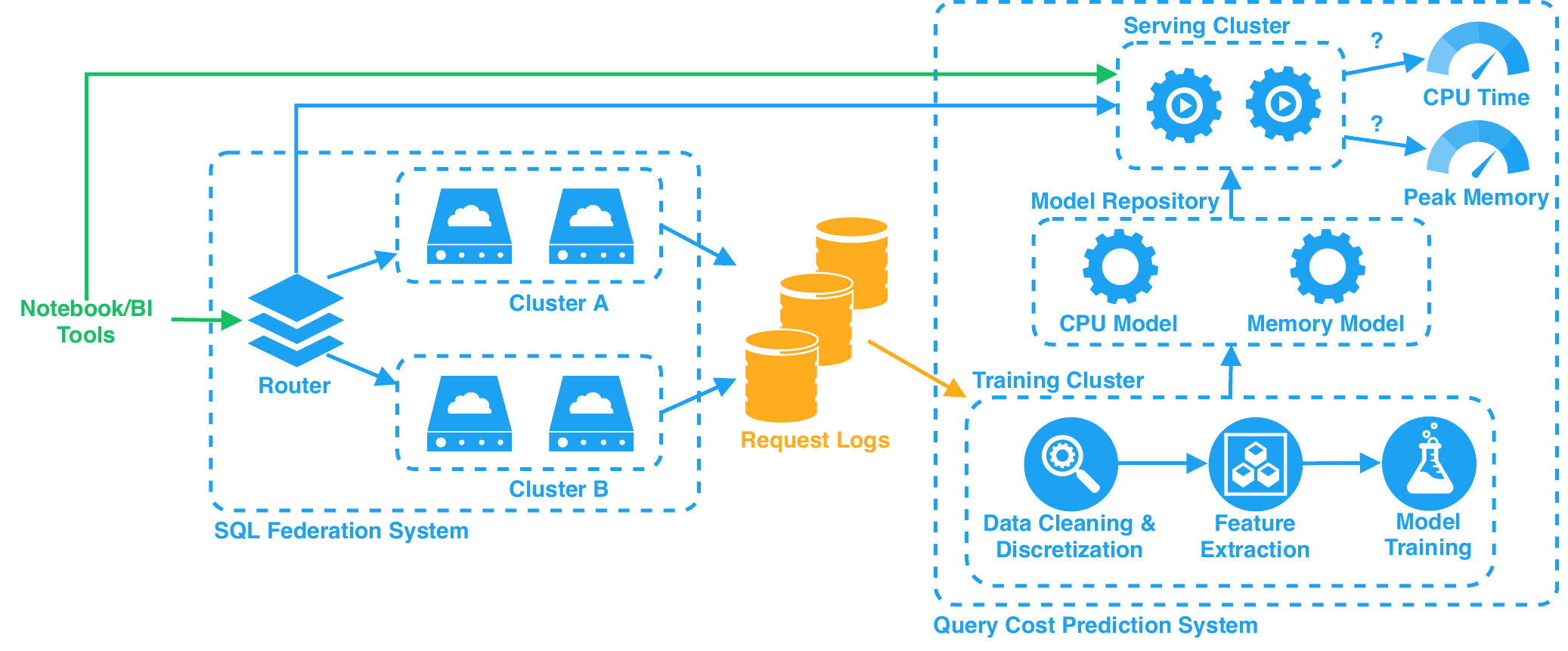}
	\caption{The architectural design of the query cost prediction system.}
	\label{fig:query-preditor}
\end{figure*}

In the SQL federation system shown in Figure \ref{fig:sql-system}, a query processing flow includes the following steps: 

\begin{enumerate}
	\item A client, such as a \textit{notebook} or \textit{BI} tool, sends a SQL query to the \textit{router}. 
	\item The router obtains the predicted resource usage of the query from the \textit{query cost predictor}.
	\item Using both the projected resource usage and the cluster statistics, the router determines a \textit{SQL engine cluster} to route the request.
	\item The router sends the cluster's URL endpoint, usually the endpoint of the coordinator node in that cluster, back to the client.
	\item The client sends the query to the specific cluster.
	\item The cluster plans the query, fetches metadata from the \textit{metadata storage}, and scans necessary datasets from the \textit{data storage}.
	\item Data results are aggregated and returned to the client.
\end{enumerate}

The SQL federation system and peripherals contain the following components:

\textbf{Notebook/Business Intelligence (BI) tools.} At Twitter, data analysts and data scientists are employing various notebook tools (e.g. Jupyter notebook and Apache Zeppelin) and BI tools (e.g. Tableau and Looker) for data analytics and visualization to gain insights from datasets. These tools send queries to the SQL federation system to get the corresponding data results.

\textbf{SQL engine clusters.} We utilize Presto as the core of each SQL engine cluster. Meanwhile, Twitter engineering is embarking on an effort to migrating ad-hoc clusters to the Google Cloud Platform (GCP), aka the ``Partly Cloudy'' \cite{partlycloudy}. Some of the SQL engine clusters have been migrated to the GCP. For now, we are maintaining a hybrid-cloud SQL federation system with both on-premises and cloud SQL engine clusters.

\textbf{Data storage.} At Twitter, the Hadoop Distributed File System (HDFS) \cite{shvachko2010hadoop} is well-acknowledged and widely used. Each SQL engine worker interacts with the HDFS for querying data, which is usually stored in an Apache Parquet \cite{parquet} format. With the Partly Cloudy strategy, the data storage has also been extended to the Google Cloud Storage (GCS). 

\textbf{Metadata storage} \cite{dal}. The metadata storage is used to provide metadata information, for example, how data files are mapped to schemas and tables.

\textbf{Router.} The router sits between the clients and SQL engine clusters, exposes a unified interface to the clients, hides cluster configuration details from the clients, and routes requests to specific clusters. To more efficiently utilize the resources on these clusters, the router also aids in balancing the workloads, with the help of the query cost predictor. Besides collecting real-time operational statistics such as P90 (90the percentile) of query latency from SQL engine clusters, the router leverages the prediction results from the query cost predictor to improve query scheduling and enable preemptive scaling.

\textbf{Query cost prediction serving cluster.} This is the serving cluster of the query cost prediction system. The predictor's job is to rapidly estimate the CPU and memory resources anticipated for a SQL query. Section \ref{sec:query-predictor-overview} discusses the architectural design of this system.

\section{Query Cost Prediction System}\label{sec:query-predictor-overview}

Figure \ref{fig:query-preditor} illustrates the architectural design of the query cost prediction system:

\textbf{Request logs.} The raw dataset is collected from query request logs stored in the HDFS. For each SQL query sent by a notebook/BI tool and processed in the SQL federation system, the query engine produces a request log. Table \ref{tab:sql-samples} shows that each SQL request log sample contains query-related information including the unique identifier, user name, environment, query statement, etc. The logs in the recent three months (90 days) are a good indicator for forecasting the cost of online queries from our experiments. Such a typical dataset consists of around 1.2 million records and more than 20 columns.

\begin{table*}[htb]
	\centering
	\caption{Samples of SQL query request logs. Sensitive information such as user names and concrete SQL statements were hydrated and replaced. Some unrelated columns have been removed from the table.}
	\label{tab:sql-samples}
	\begin{tabular}{|r|c|c|c|c|c|c|}
		\hline
		\textbf{query\_id}  & \textbf{user}    & \textbf{cluster} & \textbf{query}* & \textbf{cpu\_time\_ms}* & \textbf{peak\_memory\_bytes}* & \textbf{datehour}   \\ \hline
		Uh0u6ScxuJ & alice   & cluster\_a  & sql1  & 10143681      & 1204117281          & 2020021013 \\ \hline
		HSJb3hSEe9 & bob     & cluster\_b  & sql2  & 5903987       & 9038118972          & 2020021411 \\ \hline
		y2cysjWzKC & bob     & cluster\_a  & sql3  & 284392        & 1204117281          & 2020021719 \\ \hline
		YqtRmXL8Gy & alice   & cluster\_a  & sql4  & 53            & 45056               & 2020091516 \\ \hline
		oMawUdJuHA & charley & cluster\_a  & sql5  & 179972        & 118783230           & 2020110601 \\ \hline
		
		\multicolumn{4}{l}{*: These are the columns utilized for model training.}
	\end{tabular}
\end{table*}

\begin{figure}[htb]
	\centering
	\includegraphics[scale=0.13]{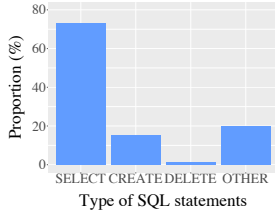}
	\caption{Distribution of SQL statements in a typical Twitter's OLAP workload based on types.}
	\label{fig:sql-type}
\end{figure}

In Twitter's data platform, most SQL statements in a typical OLAP workload are \textit{SELECT} statements, leveraged to query datasets stored in various data sources such as HDFS and GCS. Figure~\ref{fig:sql-type} illustrates the distribution of SQL statements based on types. Besides writing \textit{SELECT} statements, customers also leverage \textit{CREATE} statements to create temporary tables or material views and \textit{OTHER} statements mainly for metadata querying. \textit{DELETE} statements are rarely used.

\textbf{Training cluster.} The training cluster takes charge of the computation with machine learning techniques. We train two machine learning models-CPU model and memory model-from historical request logs for CPU time and peak memory prediction. First, we perform data cleaning and discretization to the raw dataset, converting the continuous CPU time and peak memory into buckets. Then, we apply vectorization techniques to extract features from raw SQL statements and employ classification algorithms to obtain models.

\textbf{Model repository.} The model repository manages the models, including model storage and versioning. The models are stored in a central repository such as GCS buckets. 

\textbf{Serving cluster.} The serving cluster fetches models from the model repository and wraps models into a web predictor service, exposing RESTful APIs for external usages and forecasting the CPU time and peak memory for online SQL queries from the notebook/BI tools (for customers to have a sense of the estimate of resource usages of their queries) and the router (for query scheduling and preemptive scaling).

\section{Data Preprocessing}\label{sec:preprocessing}

\subsection{Data Cleaning and Discretization}

Prior DBMS statistical approaches apply regression techniques such as time series analysis to solve DBMS problems. However, as the distribution of SQL query resource usage follows a power law, it poses challenges to conventional regression approaches such as \cite{clauset2009power} and \cite{sun2020adaptive}. Moreover, we noticed that the quick estimate of resource usages for customers and the router for scheduling and scaling do not require an accurately predicted value. For example, from our study, a customer does not care much about whether the query will cost 60 minutes or 61 minutes CPU time, but whether this is a CPU-consuming query or a lightweight query that can be completed in an acceptable range of wall clock time. For query scheduling or scaling, we also do not need an accurate value. Instead, we need to know whether a query is a low resource-consuming query, a medium one, or a high one. This judgment leads to the appliance of data discretization to the raw dataset, transforming the continuous data to discrete data. We also propose that this approach can be generalized to other DBMS problems where an accurately predicted value can be replaced by an approximate range.

\begin{figure}[htb]
	\centering
	\includegraphics[scale=0.17]{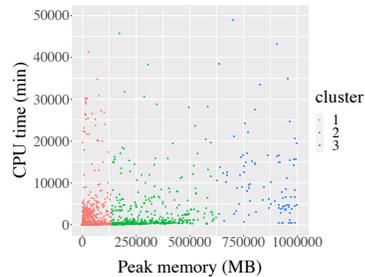}
	\caption{Clustering of 10,000 samples of peak memory and CPU time. Some samples with very high CPU time or peak memory are removed from the figure.}
	\label{fig:cpu-memory-clustering}
\end{figure}

Some prior work, such as \cite{ganapathi2009predicting} and \cite{ma2018query}, leverages clustering to predict DBMS metrics. We tried applying the k-means clustering algorithm, used in \cite{ma2018query}, to the dataset collected from request logs in 3 months. Figure \ref{fig:cpu-memory-clustering} shows an example of categorizing 10,000 queries of peak memory and CPU time in three clusters. However, we found that this clustering approach did not work well. First of all, in query scheduling and preemptive scaling, we usually plan the CPU and memory resources separately instead of collectively. Moreover, we found that the correlation between peak memory and CPU time is only 0.256, indicating a query that consumes a large amount of memory may only require a small chunk of CPU time. But the algorithm tended to cluster queries with low CPU time and those with low peak memory.

\begin{figure}[htb]
	\centering
	\begin{subfigure}[t]{0.22\textwidth}
		\centering
		\includegraphics[height=1.2in]{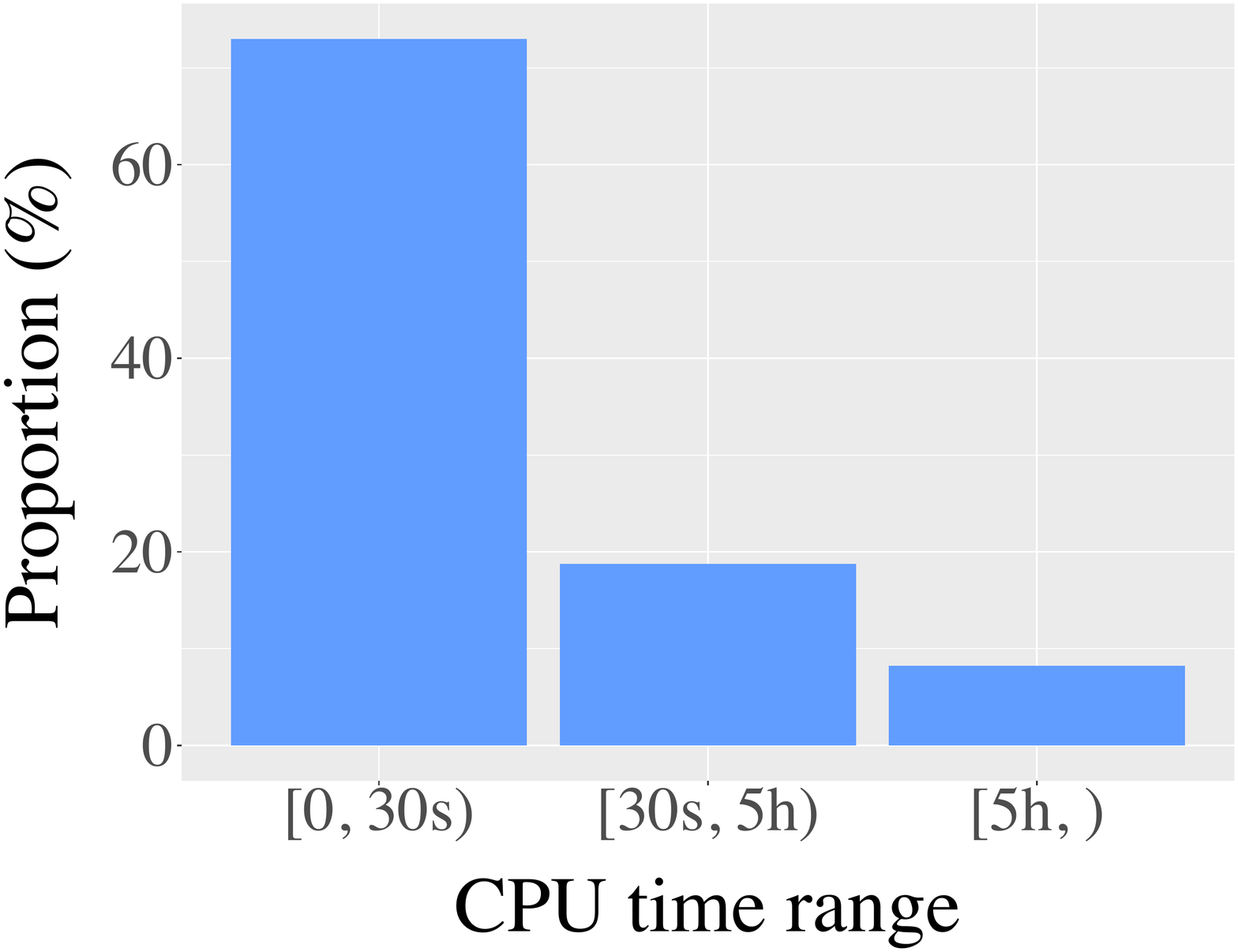}
		\caption{Distribution of queries in CPU time ranges.}
		\label{fig:cpu-dist}
	\end{subfigure}
	~
	\begin{subfigure}[t]{0.22\textwidth}
		\centering
		\includegraphics[height=1.2in]{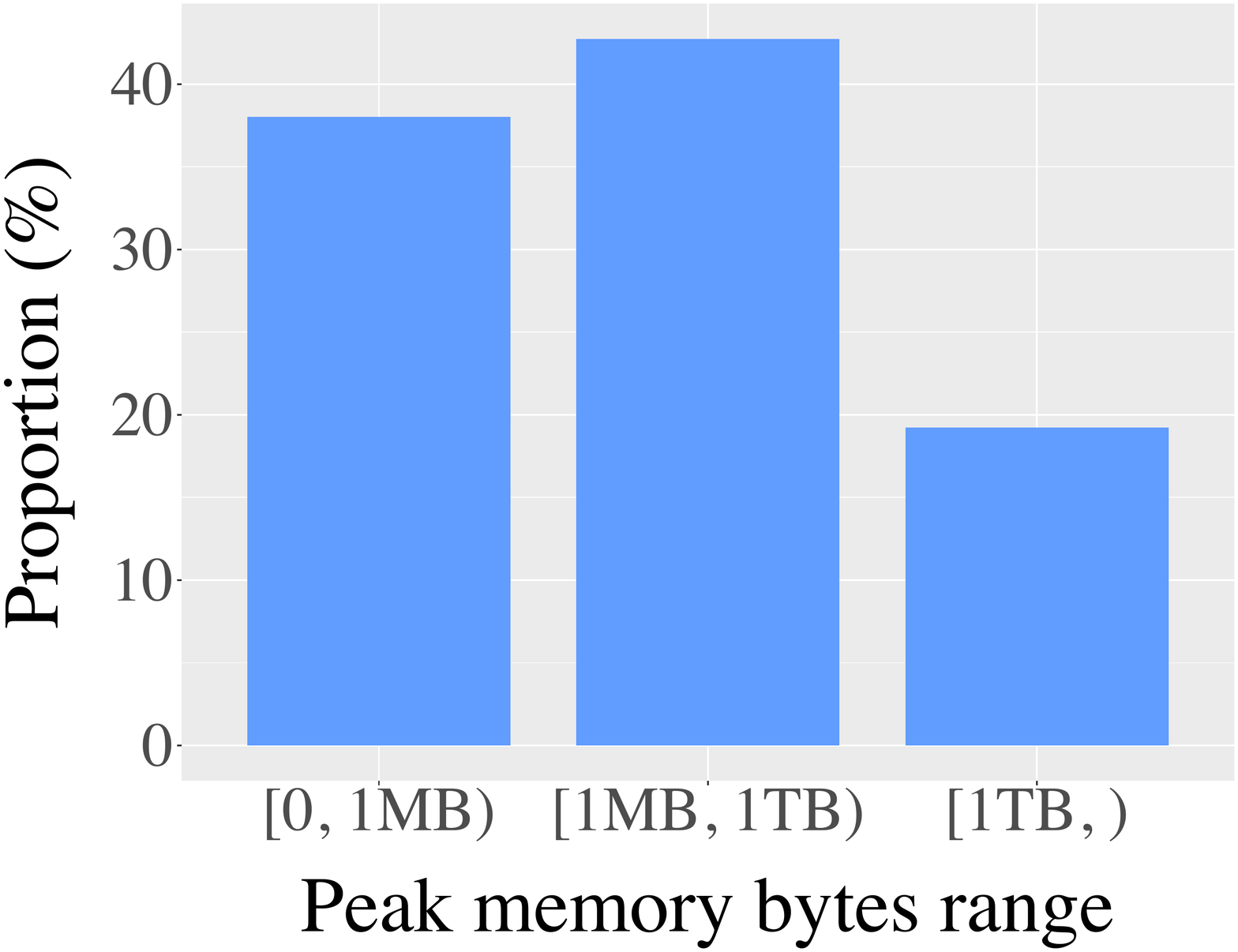}
		\caption{Distribution of queries in peak memory ranges.}
		\label{fig:memory-dist}
	\end{subfigure}
	\caption{Distribution of queries.}
	\label{fig:cpu-memory-dist}
\end{figure}

How to group these queries? First of all, we select 5h and 1TB as boundaries for high CPU usage and memory usage respectively, because these are prior thresholds applied in our SQL federation system, which are obtained from our operational experience in running analytical queries. For the CPU time, based on our DevOps experience, we consider queries whose CPU time is less than 30s as lightweight queries. This also helps us capture the large proportion of queries in the range [0, 30s), which occupies more than 70\% of queries. By contrast, only 1\% of queries fall in the range [30s, 1min). For the peak memory, we found that the distribution tends to be more evenly distributed. As a result, we generally evenly categorize the queries whose peak memory is lower than 1TB and select 1MB as the boundary for low and medium-memory-consuming queries. In summary, we categorize the CPU time into three ranges: [0, 30s), [30s, 5h), [5h, ); categorize the peak memory into three ranges: [0, 1MB), [1MB, 1TB), [1TB, ). Figure \ref{fig:cpu-memory-dist} shows the distribution of queries out of this partition.

As the queries are not evenly arranged especially for CPU time usage, this may lead to an imbalanced classes issue in the classification which we need to pay attention to. More evaluation details are discussed in Sections \ref{sec:training} and \ref{sec:serving}. After the dataset is transformed, we partition the dataset into a training dataset (80\%) and a testing dataset (20\%).

\subsection{Feature Extraction}

For the transformed dataset with peak memory and CPU time in categories, we apply vectorization techniques from Natural Language Processing (NLP) to generate essential features. Each SQL statement is mapped to a vector of numbers for subsequent processing via vectorization, making it easier for running classification algorithms on text-based data. We employ the bag-of-words (BoW) models, in which each word is represented by a number such that a SQL statement can be represented by a sequence of numbers. Word frequencies are a typical representation. The term frequency-inverse document frequency (TF-IDF) is another popular vectorization approach. BoW models are known for their high flexibility. They can be generalized to a variety of text data domains.

Word embedding is another widely-used NLP approach where words with similar meanings are represented by similar vectors, usually through computing the joint probability of words. In the word embedding, each word is mapped to a high-dimensional vector, such that various deep learning algorithms can be applied to solve NLP tasks. However, word embedding brings challenges for domain-specific context \cite{nooralahzadeh2018evaluation}, indicating it may not fit the context of Twitter SQL statements without a pre-trained domain-specific embedding model. In our work, considering there are extra training costs required for the word embedding model, whereas our BoW-based models already have high prediction accuracy and interpretability, we move forward with the BoW-based models in our machine learning pipeline.

BoW models are good at feature engineering for resource usage prediction. They produce features without computation in any SQL engines or communication with any metadata stores. We do not use table-specific statistics for feature engineering as this type of data requires additional costs for analyzing SQL statements and fetching table-related metadata. We also observed that with tree-based machine learning models where feature importance can easily be interpreted, some SQL-related information such as the access to specific tables and the limit of time ranges can be captured and reflected. This implies machine learning techniques can also help developers to gain insights from large-scale SQL systems.

\section{Model Training \& Evaluation}\label{sec:training}

After preparing the features, we trained three types of classifiers on the training dataset: Random Forest (RF), which is an ensemble learning method; XGBoost, which is a tree-based gradient boosting approach; and Logistic Regression, which predicts probabilities of certain classes with a logistic function. All of them are well-known for high computational scalability and high interpretability. They have been widely used for classification tasks. The 3-fold cross-validation is used to find the optimal hyperparameters. From our experiments, a typical training job on around 1.2 million query logs can be completed in less than 5 hours in a machine with 8 CPU cores and 64GB of memory. Then, we tested the trained classifiers on the testing dataset. As the classes are imbalanced and resource-consuming queries are of higher importance than lightweight queries, we studied the precision and recall of each class together with the overall accuracy for each of the CPU and memory models. For the feature extraction, we compared the word count approach and the TF-IDF approach. Table \ref{tab:model-accuracies} shows the comparison of these approaches.

From the evaluation results, the XGBoost model with the TF-IDF approach outperforms other CPU models up to 3\% and other memory models up to 4\% for the overall accuracy. It achieves 97.9\% accuracy for the CPU model and 97.0\% accuracy for the memory model. The Random Forest models and XGBoost models have similar performance and are higher than that of Logistic Regression models. Most TF-IDF-based models have a little bit higher accuracy than their counterparts based on word counts. 

\begin{table}[htb]
	\caption{The model accuracy(\%) of each classifier.}
	\centering
	\label{tab:model-accuracies}
	\begin{tabular}{|l|c|c|}
		\hline
		& CPU  & Memory \\ \hline
		Random Forest - Word Count       & 96.9 & 95.5   \\ \hline
		Random Forest - TF-IDF           & 97.2 & 95.4   \\ \hline
		XGBoost - Word Count             & 97.4 & 96.8   \\ \hline
		XGBoost - TF-IDF                 & 97.9 & 97.0   \\ \hline
		Logistic Regression - Word Count & 94.8 & 92.9   \\ \hline
		Logistic Regression - TF-IDF     & 95.0 & 92.9   \\ \hline
	\end{tabular}
\end{table}

Accuracy is a popular metric to evaluate model performance, but it is not the only one. In particular, considering the imbalanced classes in the training dataset, shown in Figures \ref{fig:cpu-dist} and \ref{fig:memory-dist}, a high overall accuracy does not always indicate the model is good at predicting all classes. A high capability of classifying a class containing a dominant number of samples can conceal the low accuracy of predicting classes with smaller numbers of samples. To overcome the potential issue here, we also considered the precision and recall of each class, especially the classes representing CPU or memory-intensive queries. In our work, \textit{precision} is defined as the proportion of returned results that are relevant; \textit{recall} (also known as sensitivity) is defined as the proportion of relevant cases that are returned. Details of precision and recall of the XGBoost - TD-IDF classifier are shown in Tables \ref{tab:cpu-model} and \ref{tab:memory-model}. 

\begin{table}[htb]
	\caption{The precision and recall for each class of the CPU time model.}
	\centering
	\label{tab:cpu-model}
	\begin{tabular}{|r|c|c|}
		\hline
		\textbf{CPU time}    & \textbf{Precision} & \textbf{Recall} \\ \hline
		{[}0, 30s)  & 0.98      & 0.99   \\ \hline
		{[}30s, 5h) & 0.96      & 0.95   \\ \hline
		{[}5h, )    & 0.96      & 0.95  \\ \hline
	\end{tabular}
\end{table}

\begin{table}[htb]
	\caption{The precision and recall for each class of the peak memory model.}
	\centering
	\label{tab:memory-model}
	\begin{tabular}{|r|c|c|}
		\hline
		\textbf{Peak memory}    & \textbf{Precision} & \textbf{Recall} \\ \hline
		{[}0, 1MB)  & 0.98      & 0.99   \\ \hline
		{[}1MB, 1TB) & 0.92      & 0.91   \\ \hline
		{[}1TB, )  & 0.97      & 0.98   \\ \hline
	\end{tabular}
\end{table}

From the tables, we can see that XGBoost achieves high precision and recall for all classes besides the overall accuracy. Particularly, it reaches no less than 0.95 of precision and recall for resource-consuming queries: [5h, ) and [1TB, ). XGBoost has the best performance, so we decided to use it in our online environment. Moreover, Twitter has an internal distribution of the open-sourced XGBoost JVM package, fitting Twitter's technical stack. This is also an add-on advantage for making this decision.

\section{Model Serving \& Monitoring}\label{sec:serving}

After the models are trained and tested, we encapsulate the models into a web application for real-time serving. The service, held in the serving cluster, is deployed in Aurora containers on Mesos \cite{hindman2011mesos}, which is the stack widely used at Twitter. As each deployment unit is stateless, the application's scalability can be enhanced by increasing the number of deployment replicas. The models are stored in a central repository and fetched by the application when a deployment unit starts. The service exposes two RESTful API endpoints to forecast CPU time and peak memory of a SQL query. The inference time is around 200ms. 

When the models serve online requests, concept drift is observed since unseen features emerge over time, deteriorating the model performance. Hence, model monitoring and retraining are required under a given time granularity. From our DevOps experience of the SQL federation system, the query counts have high variance daily but are quite stable weekly. For example, we usually observe many more queries on Mondays than those on Fridays. Furthermore, a large number of queries are scheduled to run weekly. To reduce the influence of outliers or peaks in some specific workdays, we re-evaluate the online models weekly.

\begin{figure}[htb]
	\centering
	\begin{subfigure}[t]{0.3\textwidth}
		\centering
		\includegraphics[scale=0.14]{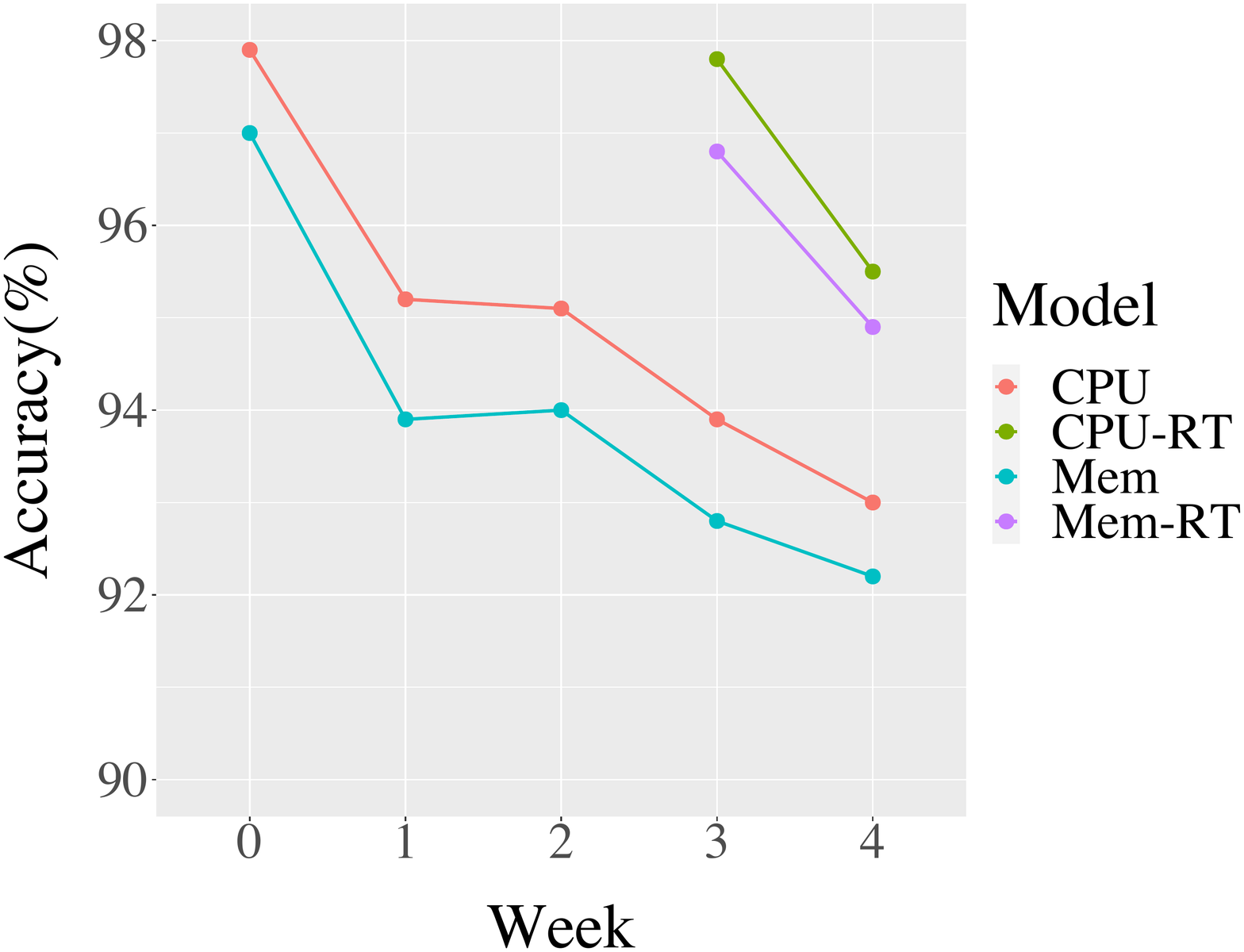}
		\caption{Changes of model \textbf{accuracy}.}
		\label{fig:model-monitoring-accuracy}
	\end{subfigure}
	~\\
	\begin{subfigure}[t]{0.22\textwidth}
		\centering
		\includegraphics[scale=0.14]{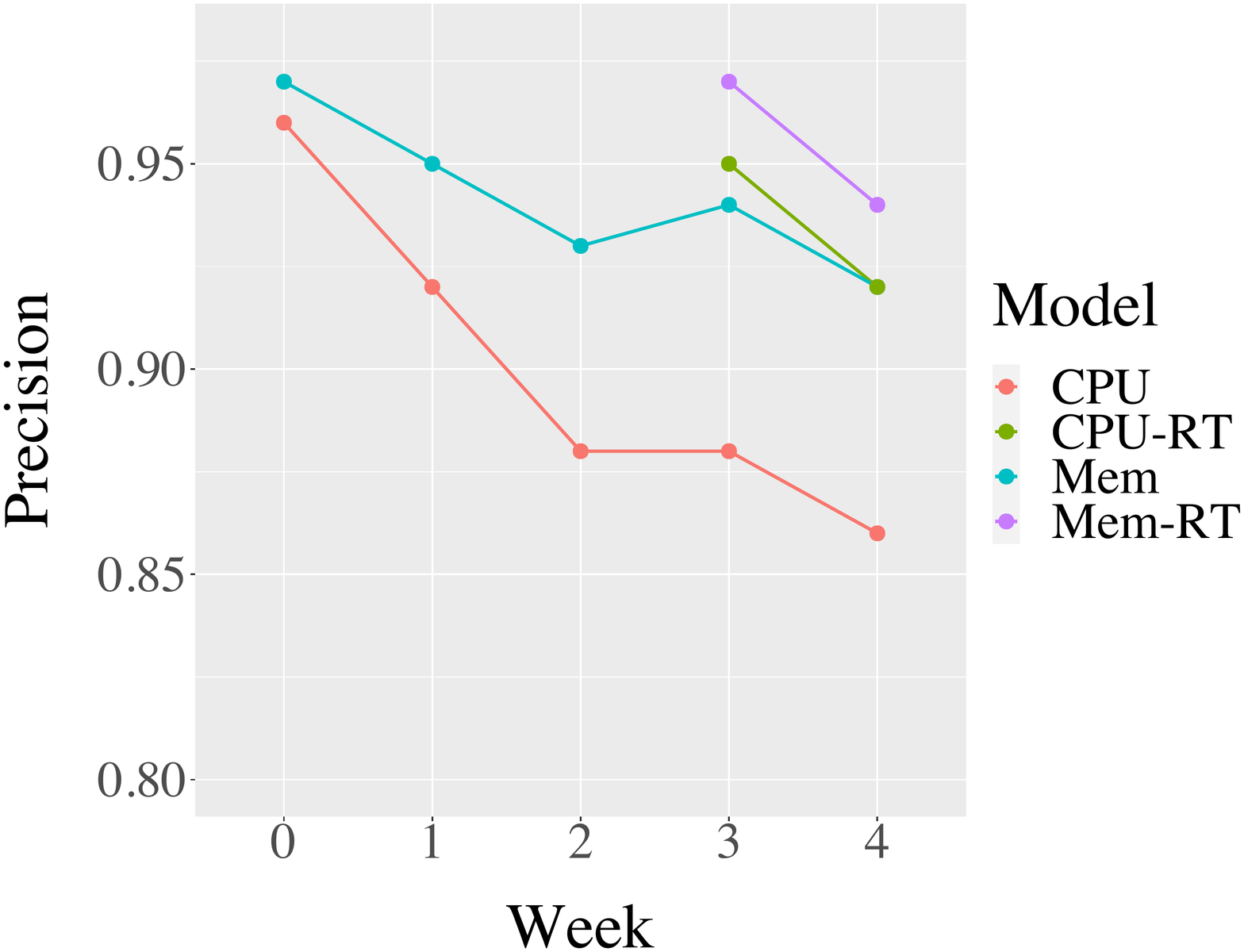}
		\caption{Changes of model \textbf{precision} for [5h, ) and [1TB, ).}
		\label{fig:model-monitoring-precision}
	\end{subfigure}
	~
	\begin{subfigure}[t]{0.22\textwidth}
		\centering
		\includegraphics[scale=0.14]{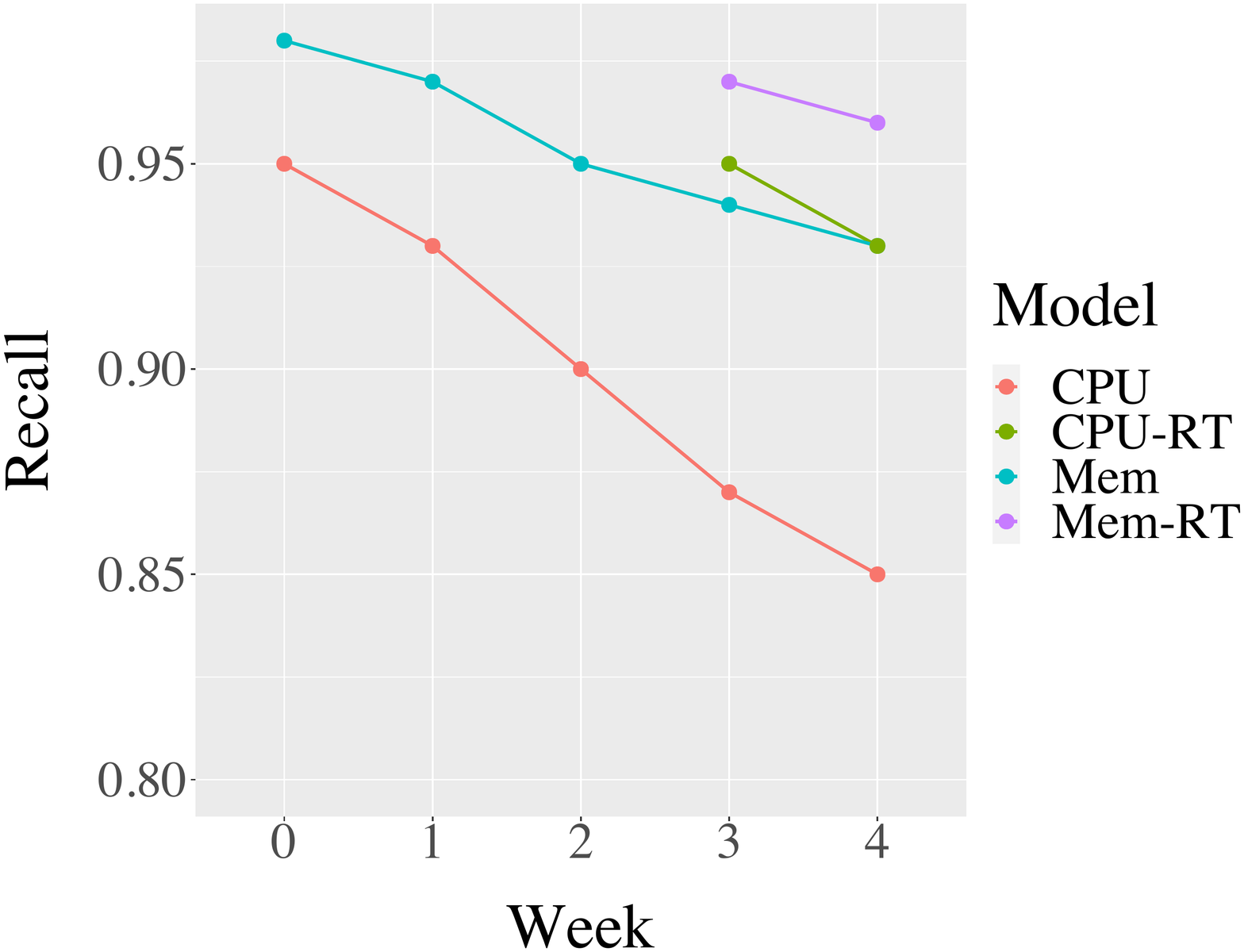}
		\caption{Changes of model \textbf{recall} for [5h, ) and [1TB, ).}
		\label{fig:model-monitoring-recall}
	\end{subfigure}
	\caption{Changes of model performance over time. RT: Retrained models.}
	\label{fig:model-monitoring}
\end{figure}

Figure \ref{fig:model-monitoring} illustrates the drifts of models in a real-time online environment. Because resource-consuming queries weigh more than lightweight queries, we also include the precision and recall for resource-consuming ranges: [5h, ) and [1TB, ). As time goes, the model metrics are gradually decreasing. For example, at first, the overall accuracy of the CPU model is as high as around 98\%. After four weeks, the accuracy shrank to around 93\%. The recall of the CPU time class [5h, ) has dropped from 0.96 to 0.85. This is usually caused by unseen features in the latest queries which are not captured by the models we have trained. To mitigate the concept drift, we can retrain the models when both the precision and recall for CPU or memory-intensive queries are lower than 0.9. In Figure \ref{fig:model-monitoring}, at week 3, the precision of the class [5h, ) is 0.88 and the recall is 0.87. As both precision and recall are less than 0.9, this will trigger retraining of the CPU model. To maintain the consistency of training timestamps of the CPU and memory models, we will also retrain and redeploy the memory model. After the model retraining with the same hyperparameters, the metrics will go back to a high level. This indicates our models have a high adaptivity of new features.

We also observed that even though the CPU model accuracy is higher than that of the memory model, the precision and recall are lower and fall faster. For example, the precision of both the CPU range [5h, ) and the memory range[1TB, ) is higher than 0.95 initially. But after four weeks, the precision of the CPU range has decayed to 0.86. By contrast, the precision of the memory range is still 0.92. This is probably because the setting of the CPU range [5h, ) is too high, causing the model to fail to capture the features of CPU-consuming queries. As a lesson learned from this study, we are considering tuning the CPU range for this type of queries in the future.

\section{Conclusion}\label{sec.conclusion}

In this paper, we introduced the SQL federation system in the Twitter data platform and proposed a SQL query cost prediction system. Unlike prior work, the proposed system learns from plain SQL statements, builds machine learning models from historical query request logs, and forecasts CPU time and peak memory ranges. The evaluation shows that the proposed system of the XGBoost classifier with TF-IDF vectorization can achieve 97.9\% accuracy for CPU time prediction and 97\% accuracy for memory usage prediction. We deployed the models in the production environment and implemented a quantitative analysis on the concept drift observed. This work can help provide an estimate of resource usage to customers, improve query scheduling, and enable preemptive scaling, without dependency on SQL engines or query plans.

\section*{Acknowledgment}

We would like to express our gratitude to everyone who has served on Twitter's Interactive Query team, including former team members Hao Luo and Yaliang Wang. We are also grateful to Daniel Lipkin and Derek Lyon for their strategic vision, direction, and support to the team. Finally, we thank Alyson Pavela, Julian Moore, and the anonymous IC2E reviewers for their insightful suggestions that helped us significantly improve this paper.

\Urlmuskip=0mu plus 1mu\relax  
\bibliographystyle{IEEEtran}
\bibliography{library}

\end{document}